\documentclass{article}

\usepackage{arxiv}

\usepackage[dvipsnames]{xcolor}

\usepackage[utf8]{inputenc} % allow utf-8 input
\usepackage[T1]{fontenc}    % use 8-bit T1 fonts
\usepackage{hyperref}       % hyperlinks
\usepackage{url}            % simple URL typesetting
\usepackage{booktabs}       % professional-quality tables
\usepackage{amsfonts}       % blackboard math symbols
\usepackage{nicefrac}       % compact symbols for 1/2, etc.
\usepackage{microtype}      % microtypography
\usepackage{lipsum}
\usepackage{multirow, amsmath, graphicx, multirow, tabularx, threeparttable, subcaption}
\usepackage{diagbox}
\usepackage[abbreviations]{glossaries-extra}
\glssetcategoryattribute{acronym}{indexonlyfirst}{true}
\newabbreviation{svm}{SVM}{support vector machine}
\newabbreviation{mae}{MAE}{mean absolute error}
\newabbreviation{rmse}{RMSE}{root mean square error}
\def\JRcomment[#1]{\textcolor{Red}{#1}}
\def\MCcomment[#1]{\textcolor{Blue}{#1}}

\title{Guitar Effects Recognition and Parameter Estimation with Convolutional Neural Networks}

\author{
  Marco Comunità, Dan Stowell, Joshua D. Reiss\\
  \texttt{\{m.comunita, dan.stowell, joshua.reiss\}@qmul.ac.uk}\\\\
  Centre for Digital Music\\
  Queen Mary University of London\\
  UK \\
}

\begin{document}
\maketitle

%===================================================%
%   ABSTRACT
%===================================================%
\begin{abstract}
Despite the popularity of guitar effects, there is very little existing research on classification and parameter estimation of specific plugins or effect units from guitar recordings. In this paper, convolutional neural networks were used for classification and parameter estimation for 13 overdrive, distortion and fuzz guitar effects. A novel dataset of processed electric guitar samples was assembled, with four sub-datasets consisting of monophonic or polyphonic samples and discrete or continuous settings values, for a total of about 250 hours of processed samples. Results were compared for networks trained and tested on the same or on a different sub-dataset. We found that discrete datasets could lead to equally high performance as continuous ones, whilst being easier to design, analyse and modify. Classification accuracy was above 80\%, with confusion matrices reflecting similarities in the effects timbre and circuits design. With parameter values between 0.0 and 1.0, the mean absolute error is in most cases below 0.05, while the root mean square error is below 0.1 in all cases but one.
\end{abstract}

% keywords can be removed
\keywords{Guitar effects classification \and Guitar effects parameters \and Audio effects \and Deep learning}

%===================================================%
%   INTRODUCTION
%===================================================%
\section{Introduction}
In music composition, production and engineering, audio effects play an essential role in altering the sound towards the desired final result. For instruments like the electric guitar, the processing signal chain can often be viewed as part of the artist’s creative expression \cite{case2010recording}.
Entire musical genres and styles are frequently defined and identified by the type of audio effects adopted \cite{blair2015southern, williams2012tubby}; and renowned musicians commonly rely on specific combinations of guitars, amplifiers and effects to achieve a unique sound \cite{prown2003gear}. 
Through the decades, artists, engineers and producers have defined a palette of sounds that have become a reference for guitar players. In the effort to recreate a specific sound or atmosphere, professionals and amateurs go to great lengths to identify the exact gear that was used in a certain recording.
When describing a desired result, people often rely on naming a reference style, artist or song, rather than talking in terms of sound features or effect parameters.

Although the design, reconstruction and emulation of audio effects is well-studied \cite{zolzer2011dafx, pakarinen2009review}; it is less so for their recognition and their parameter estimation. Therefore, in our work, we set out to develop a deep learning model capable of recognising which specific guitar pedal effect was used in a recording as well as estimating its parameters. Being the single most important effect for electric guitar, and the one that usually triggers most discussions, this work focused on nonlinear effects, i.e. overdrive, distortion and fuzz.

The rest of this paper is organised as follows: in section \ref{sec:background} we look at the state of the art and work related to guitar effects recognition, section \ref{sec:dataset} introduces the dataset we assembled for this work, while the networks architecture is described in section \ref{sec:architecture}; sections \ref{sec:experiments}, \ref{sec:results} and \ref{sec:conclusion} outline experiments, results and conclusions.

%===================================================%
%   BACKGROUND
%===================================================%
\section{Background}
\label{sec:background}
The recognition of musical instrument sounds has been of interest to the information retrieval community for long time \cite{herrera2003automatic}; with applications in musical sounds databases, intelligent music search and recommendation systems. The estimation of instruments' parameters, as well as the classification of playing gestures or styles has also been extensively studied - and applied in contexts like automatic music transcription \cite{benetos2013automatic, kehling2014automatic}, music education \cite{dittmar2012music} or musicology \cite{abesser2016score}.

Many papers focused on guitar parameters. String, fret and plucking position estimation have been extensively studied, including works on classical and acoustic \cite{barbancho2009pitch} and electric \cite{abesser2012automatic} guitar. 
Most works are based on features extraction from recorded sounds \cite{barbancho2009pitch, abesser2012automatic, dittmar2013real, kehling2014automatic, geib2017automatic}, but there are also examples of estimation based on guitar-string physical models \cite{hjerrild2019estimation, hjerrild2019physical} for high-tempo and real-time applications.
Plucking and pickup position estimation has also been studied, with Mohamad et al exploring solutions based on spectral features (comparing recordings with string models) and autocorrelation \cite{mohamad2017pickup2}. The study was also extended to the case of nonlinear audio effects in the signal chain \cite{mohamad2017pickup}.

Classification of playing styles and techniques has also received substantial attention. In \cite{abesser2010feature}, the authors compare the performance of several classifiers (\gls{svm}, Gaussian mixture models, nearest neighbours) on 5 plucking styles (e.g. finger, pick, slap) from bass guitar recordings. In \cite{schuller2015parameter} Schuller et al extend the work to include expression styles (e.g. bend, slide, vibrato). Su et al, in \cite{su2014sparse}, apply sparse dictionary learning to classify guitar playing techniques (e.g. vibrato, hammer-on, pull-off). The same classification problem is solved in \cite{wang2020spectral} using a deep belief network.

Other examples of work on guitar-related classification problems focused on playing mode (e.g. bass, solo melodic improvisation, chordal playing) \cite{foulon2013automatic}, chords fingering \cite{barbancho2011automatic} and guitar model \cite{dosenbach2008identification, johnson2015guitar, profeta2019feature, profeta2019comparison}. 

However, there is only a small corpus of research on guitar effects recognition \cite{stein2010automatic, stein2010automatic2, eichas2015feature, schmitt2017recognising}.

In \cite{stein2010automatic}, Stein worked with guitar and bass recordings on recognition of 11 different effects: feedback and slapback delay, reverb, chorus, flanger, phaser, tremolo, vibrato, distortion and overdrive. In a subsequent work \cite{stein2010automatic2}, the author extended his method - based on spectral, cepstral and harmonic features and \gls{svm}s - to cascaded effects. 
In \cite{schmitt2017recognising}, using the same dataset, the authors aimed to understand which are the most relevant features for the classification task adopting a ”bag-of-audio-words” approach; while in \cite{eichas2015feature}, the authors - making use of specific input test signals, and extracting features from the output - worked on classification of 10 analogue effect units into 5 categories. In \cite{eichas2018gray}, the guitar amplifier modelling process includes emulation of nonlinear blocks and estimation of parameters. In these last examples, the approach is limited to the case in which the unit to classify/model is accessible, which defeats the purpose of estimation from recordings.

The closest study to our work is \cite{jurgens2020recognizing}, and it is the only one that estimates effects parameters from guitar recordings. Similarly to the previous studies, the authors used SVMs to classify the same effects listed above with a reduced features set; and extended the work by training shallow neural networks on parameter estimations for 3 effects (distortion, tremolo, delay). The main limitation of this study is the necessity of separate features selection and network training for each effect.

In all these cases, the authors worked on generic audio effects or categories and, to the best of our knowledge, there is no previous research focusing on classification and parameter estimation of specific plugins or effect units from guitar recordings.

%===================================================%
%   DATASET
%===================================================%
\section{Dataset}
\label{sec:dataset}
We assembled a novel dataset of processed electric guitar samples using unprocessed recordings from the IDMT-SMT-Audio-Effects dataset \cite{stein2010automatic}. The dataset \footnote{\href{https://www.idmt.fraunhofer.de/en/business_units/m2d/smt/audio_effects.html}{https://www.idmt.fraunhofer.de/en/business\_units/m2d/smt/audio\_effects.html}} includes monophonic (624 single notes) and polyphonic (420 intervals and chords) recordings (wav - 44.1kHz, 16bit, mono) from 2 different electric guitars, each with two pick-up settings and up to 3 plucking styles. The monophonic recordings cover the common pitch range of a 6-string electric guitar, and the polyphonic samples were obtained mixing single notes recordings to generate two-notes intervals and 3- or 4-notes chords. All samples are 2 seconds long. The monophonic recordings required removal of background noise before the note onset, which we obtained using a python script together with \textit{Librosa}'s \cite{mcfee2015librosa} onset detection function.

To assemble our dataset we selected 13 overdrive, distortion and fuzz plug-ins (see Table \ref{table:plugins}) designed to emulate some of the most iconic and widely used analogue guitar effect pedals. All the plugins have 2 or 3 controls and, regardless of the specific name adopted by the designer, the controls can be identified by their processing function: Level, Gain, Tone/Equalisation. 

For training and testing purposes, 4 sub-datasets were generated, which will be referred to as Mono Discrete, Poly Discrete, Mono Continuous and Poly Continuous. 

The first two subsets (Mono Discrete, Poly Discrete) use a discrete set of combinations selected as the most common and representative settings a person might use: Gain = [0.0, 0.1, 0.2, 0.5, 0.8, 1.0], Tone/Eq = [0.0, 0.2, 0.5, 0.8, 1.0]. Also, since the Level control has no effect on the output timbre it was set to 1.0 for every combination. A summary of the controls and settings is shown in Table \ref{table:settings}. Most plugins do not include Gain values below 0.2 - this is because for such values there is no audible change between input and output or even a level attenuation (with no output for Gain = 0.0). Every monophonic and polyphonic sample was processed with all the combinations, generating a total of ${\sim}$200000 processed samples (${\sim}$120000 monophonic, ${\sim}$80000 polyphonic), for a total of about 110 hours.

For the second two subsets (Mono Continuous, Poly Continuous), both unprocessed samples as well as settings' values were drawn from a uniform distribution. We generated 10000 random samples for each plugin, obtaining a total of 260000 samples (130000 monophonic, 130000 polyphonic), equivalent to about 140 hours. Settings values were limited to fall between the extremes shown in Table \ref{table:settings}.

Generating these four subsets we aimed at gaining a deeper understanding about the generalisation capabilities of our models. The samples' were processed in MATLAB - making use of its VST plugin host features - and both unprocessed inputs and processed outputs were normalised to -6dBFS.

\begin{table}
\small
\centering
\setlength{\tabcolsep}{4pt}
\begin{tabular}{|c|c|c|c|}
\hline
\textbf{Designer} & \textbf{Plugin} & \textbf{Emulation of} & \textbf{Id}\\
\hline
\multirow{10}{*}{Audified}   
    &   
        &   Ibanez TS808
            &   808\\
    &   
        &   Ibanez TS9
            &   TS9\\
    &   
        &   Boss BD2
            &   BD2\\
    &   
        &   Boss OD1
            &   OD1\\
    &   MultiDrive
        &   Boss SD1
            &   SD1\\
    &   Pedal Pro
        &   Boss DS1
            &   DS1\\
    &   
        &   ProCo Rat
            &   RAT\\
    &   
        &   MXR Distortion+
            &   DPL\\
    &   
        &   Arbiter Fuzz Face
            &   FFC\\
    &   
        &   EH Big Muff
            &   BMF\\
\hline
\multirow{1}{*}{Mercuriall}
                &   Greed Smasher   
                &   Mesa/Boogie Grid Slammer
                &   MGS\\
\hline
\multirow{2}{*}{\shortstack{Analog\\Obsession}}
                &   Pig Pie   
                &   EH Russian Big Muff
                &   RBM\\
                &   Zupaa   
                &   Vox Tone Bender
                &   VTB\\
\hline
\end{tabular}
\caption{plugins}
\label{table:plugins}
\end{table}

\begingroup
\setlength{\tabcolsep}{4pt} % Default value: 6pt
\renewcommand{\arraystretch}{0.7} % Default value: 1
\begin{table}
    \small
    \centering
    \setlength{\tabcolsep}{4pt}
    \begin{tabular}{|c|c|c|c|}
        \hline
        \textbf{Id}
            & \textbf{Level}
                & \textbf{Gain}
                    & \textbf{Tone/Eq}\\
        \hline
        808
            & [1.0]
                & [0.2, 0.5, 0.8, 1.0]
                    & [0.0, 0.2, 0.5, 0.8, 1.0]\\
        TS9
            & [1.0]
                & [0.2, 0.5, 0.8, 1.0]
                    & [0.0, 0.2, 0.5, 0.8, 1.0]\\
        BD2
            & [1.0]
                & [0.2, 0.5, 0.8, 1.0]
                    & [0.0, 0.2, 0.5, 0.8, 1.0]\\
        OD1
            & [1.0]
                & [0.2, 0.5, 0.8, 1.0]
                    & ---\\
        SD1
            & [1.0]
                & [0.2, 0.5, 0.8, 1.0]
                    & [0.0, 0.2, 0.5, 0.8, 1.0]\\
        DS1
            & [1.0]
                & [0.2, 0.5, 0.8, 1.0]
                    & [0.0, 0.2, 0.5, 0.8, 1.0]\\
        RAT
            & [1.0]
                & [0.2, 0.5, 0.8, 1.0]
                    & [0.0, 0.2, 0.5, 0.8, 1.0]\\
        DPL
            & [1.0]
                & [0.2, 0.5, 0.8, 1.0]
                    & ---\\
        FFC
            & [1.0]
                & [0.0, 0.2, 0.5, 0.8, 1.0]
                    & ---\\
        BMF
            & [1.0]
                & [0.2, 0.5, 0.8, 1.0]
                    & [0.0, 0.2, 0.5, 0.8, 1.0]\\
        MGS
            & [1.0]
                & [0.2, 0.5, 0.8, 1.0]
                    & [0.0, 0.2, 0.5, 0.8, 1.0]\\
        RBM
            & [1.0]
                & [0.2, 0.5, 0.8, 1.0]
                    & [0.0, 0.2, 0.5, 0.8, 1.0]\\
        VTB
            & [1.0]
                & [0.1, 0.2, 0.5, 0.8, 1.0]
                    & ---\\
        \hline
    \end{tabular}
    \caption{settings}
    \label{table:settings}
\end{table}
\endgroup

%===================================================%
%   ARCHITECTURE
%===================================================%
\section{Architecture}
\label{sec:architecture}
Our neural network architecture (Table \ref{table:architecture}) is based on a combination of 2 convolutional and 3 fully connected layers, with batch normalisation layers at each hidden level. Except for the output layers' size and activation functions, the same configuration was used to train networks on both effects classification and settings estimation. Four different networks - and training paradigms - were implemented: 
\begin{itemize}
    \item effects classification network (FxNet)
    \item settings estimation network (SetNet)
    \item multitask classification and estimation network (MultiNet) - where the 2 convolutional layers are shared.
    \item settings estimation conditional network (SetNetCond) - where an extra embedding layer is added to condition the estimation on the effect class.
\end{itemize}
The loss functions adopted for the classification and estimation problems were, respectively, the cross-entropy loss and the mean square error (MSE). To evaluate the settings estimation networks, we also defined an accuracy metric for which a prediction is considered correct when the absolute error for every parameter is less then 0.1. For effects that do not have a Tone/Eq control, we represented the absence using a value of -1.0 for the prediction. The threshold of 0.1 was chosen to simplify the comparison of networks performance in different training settings and on different datasets. The value is based on the authors' experience and informal listening tests during the dataset creation phase. However, parameters' sensitivity varies between effects and controls, and differences in absolute value often do not relate linearly with perceptual differences. To overcome this limitations we do rely also on \gls{mae} and \gls{rmse} to evaluate our models.

As input features to all our networks we used mel power-spectrograms, extracted from audio in 23 ms Hann windows with 50\% overlap. In total, 128 mel-bands were used in the 0–22,050 Hz range. For a given 2 s audio input, the feature extraction produced a T x 128 output (T = 87).
These features are motivated by human auditory perception, and are a common choice in acoustic scene classification \cite{abesser2020review}. Due to the good performance of our models, we did not deem important for this study to test other features, but it could be worth exploring in the future.

% (Table \ref{table:architecture})
\begin{table}
    \small
    \centering
    \begin{threeparttable}
        \begin{tabular}{llll}
        \hline\hline
        \textbf{Layer} 
            & \textbf{Size} 
                & \textbf{\#Fmaps} 
                    & \textbf{Activation}\\
        \hline
        Convolution 2D 
            & 5x5 
                & 6 
                    & Linear\\
        Batch Normalisation 
            & - 
                & - 
                    & -\\
        Activation 
            & - 
                & - 
                    & ReLU\\
        Max Pooling 
            & 2x2 
                & -     
                    & -\\
        Convolution 2D 
            & 5x5 
                & 12 
                    & Linear\\
        Batch Normalisation 
            & - 
                & - 
                    & -\\
        Activation 
            & - 
                & - 
                    & ReLU\\
        Max Pooling 
            & 2x2 
                & -     
                    & -\\
        Fully Connected
            & 120
                & -
                    & Linear\\
        Batch Normalisation 
            & - 
                & - 
                    & -\\
        Activation 
            & - 
                & - 
                    & ReLU\\
        Fully Connected
            & 60
                & -
                    & Linear\\
        Batch Normalisation 
            & - 
                & - 
                    & -\\
        Activation 
            & - 
                & - 
                    & ReLU\\
        Fully Connected
            & (1)
                & -
                    & (2)\\
        \hline
        \end{tabular}
    \begin{tablenotes}
        \item[(1)] FxNet = \#Plug-ins - SetNet = \#Settings
        \item[(2)] FxNet = Linear - SetNet = Tanh\\
    \end{tablenotes}
   
    \end{threeparttable}
    \caption{architecture}
    \label{table:architecture}
\end{table}

%===================================================%
%   EXPERIMENTS
%===================================================%
\section{Experiments}
\label{sec:experiments}
Some preliminary experiments were conducted to obtain baseline performances for the settings estimation problem and to compare the results when using a multitask approach or a conditional network. The literature shows how multitask learning can be effective at solving related tasks \cite{ruder2017overview} and more efficient than training several networks. In the multitask paradigm, the network was trained to classify a sample and estimate its settings at the same time. In the conditional paradigm the networks were trained in "series", with the classification network (FxNet) used to condition the settings estimation network (SetNetCond). The experiments were conducted on the Mono Discrete dataset.

All networks were trained for 50 epochs, which resulted in the following test accuracy:
\begin{itemize}
    \item SetNet = 40.3\%
    \item MultiNet = 40.88\% (87.0\% classification accuracy, 44.6\% estimation accuracy)
    \item FxNet + SetNetCond = 57.3\% (89.7\% classification accuracy, 60.7\% estimation accuracy)
\end{itemize}
The results show no appreciable difference in effect classification accuracy between the multitask and the conditional paradigms but do show an impact on the settings estimation accuracy, with the conditional network performing better. Further experiments were therefore centred on the analysis of the classification network (FxNet) and the conditional estimation network (SetNetCond) when trained/tested on the 4 different sub-datasets. In the following section we illustrate the results of training the networks for 100 epochs with an early stop when the validation accuracy sees no improvement for 15 epochs. For weights update we opted for the Adam optimiser \cite{kingma2014adam} with a fixed learning rate of 0.001.

%===================================================%
%   RESULTS
%===================================================%
\section{Results}
\label{sec:results}
%===================================================%
%   RESULTS: FX CLASSIFICATION
%===================================================%
\subsection{Effects Classification}
Table \ref{table:fxnet_acc} shows the accuracy results for the effect classification problem. Note that the test accuracy is higher for networks trained on continuous datasets, respectively 90.9\% for the Mono Continuous dataset and 91.4\% for the Poly Continuous dataset. At the same time, networks trained on discrete datasets performed better when tested on continuous ones than the opposite condition (networks trained on continuous and tested on discrete datasets): 83.1\% vs 81.3\% for monophonic samples and 89.4\% vs 84.1\% for polyphonic ones.

\begingroup
\setlength{\tabcolsep}{4pt} % Default value: 6pt
\renewcommand{\arraystretch}{1} % Default value: 1
\begin{table}[h]
    \centering
    \small
    \begin{tabular}{|c|c|c|c|c|}
    \hline
        \backslashbox[]{Train}{Test}
            & \textbf{Mono Disc.} 
                & \textbf{Mono Cont.} 
                    & \textbf{Poly Disc.} 
                        & \textbf{Poly Cont.}\\
        \hline
        \textbf{Mono Disc.}
            &  86.3
                & 83.1
                    & ---
                        & ---\\
        \textbf{Mono Cont.}
            &  81.3
                & 90.9
                    & ---
                        & ---\\
        \textbf{Poly Disc.}
            & ---
                & ---
                    & 88.4
                        & 89.4\\
        \textbf{Poly Cont.}
            & ---
                & ---
                    & 84.1
                        & 91.4\\
        \hline
    \end{tabular}
    \caption{FxNet accuracy (\%)}
    \label{table:fxnet_acc}
\end{table}
\endgroup

\begin{figure*}
    \hspace{-0.4cm}
    \subfloat{\includegraphics[width=.515\linewidth]{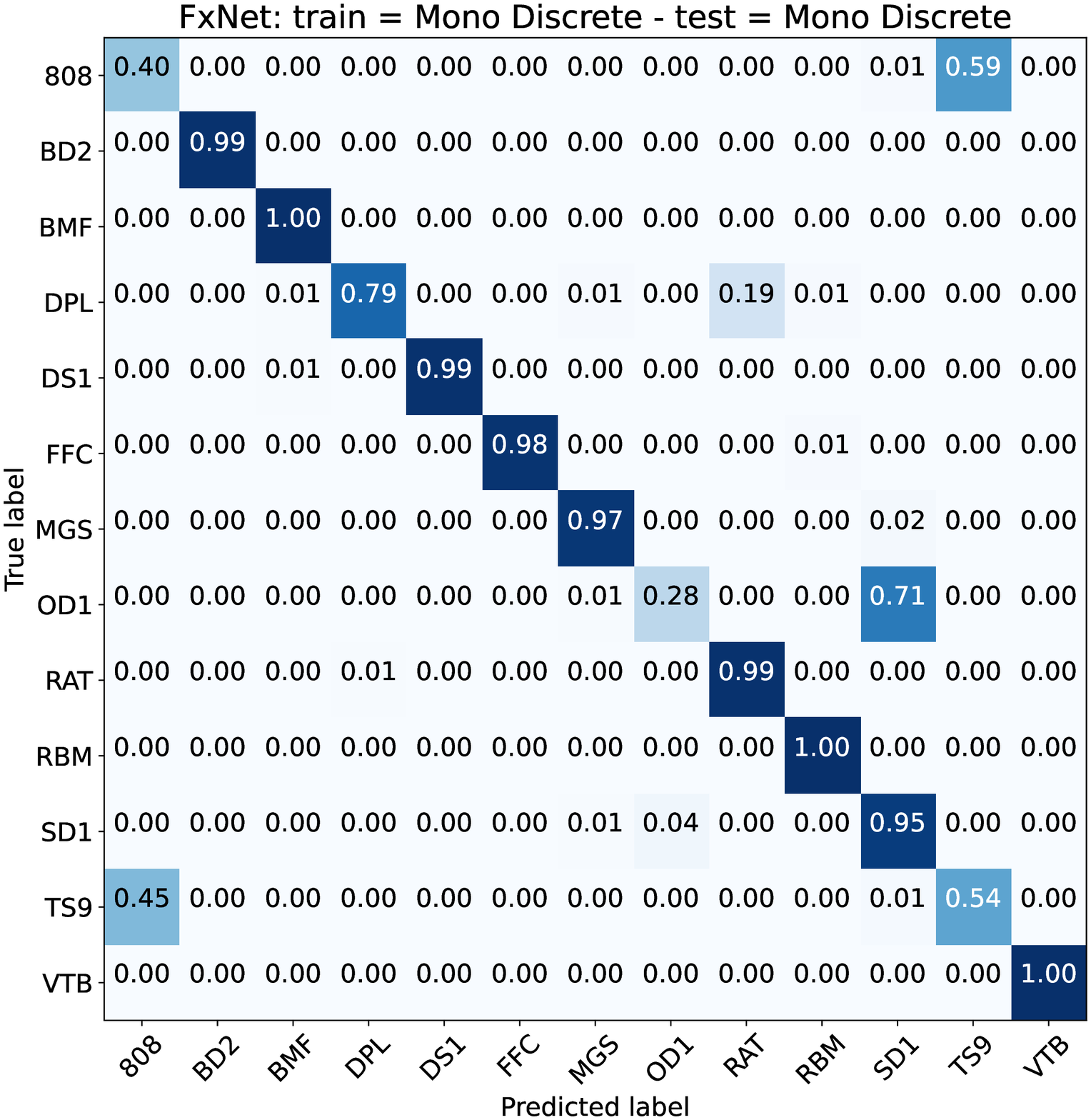}}
    \subfloat{\includegraphics[width=.515\linewidth]{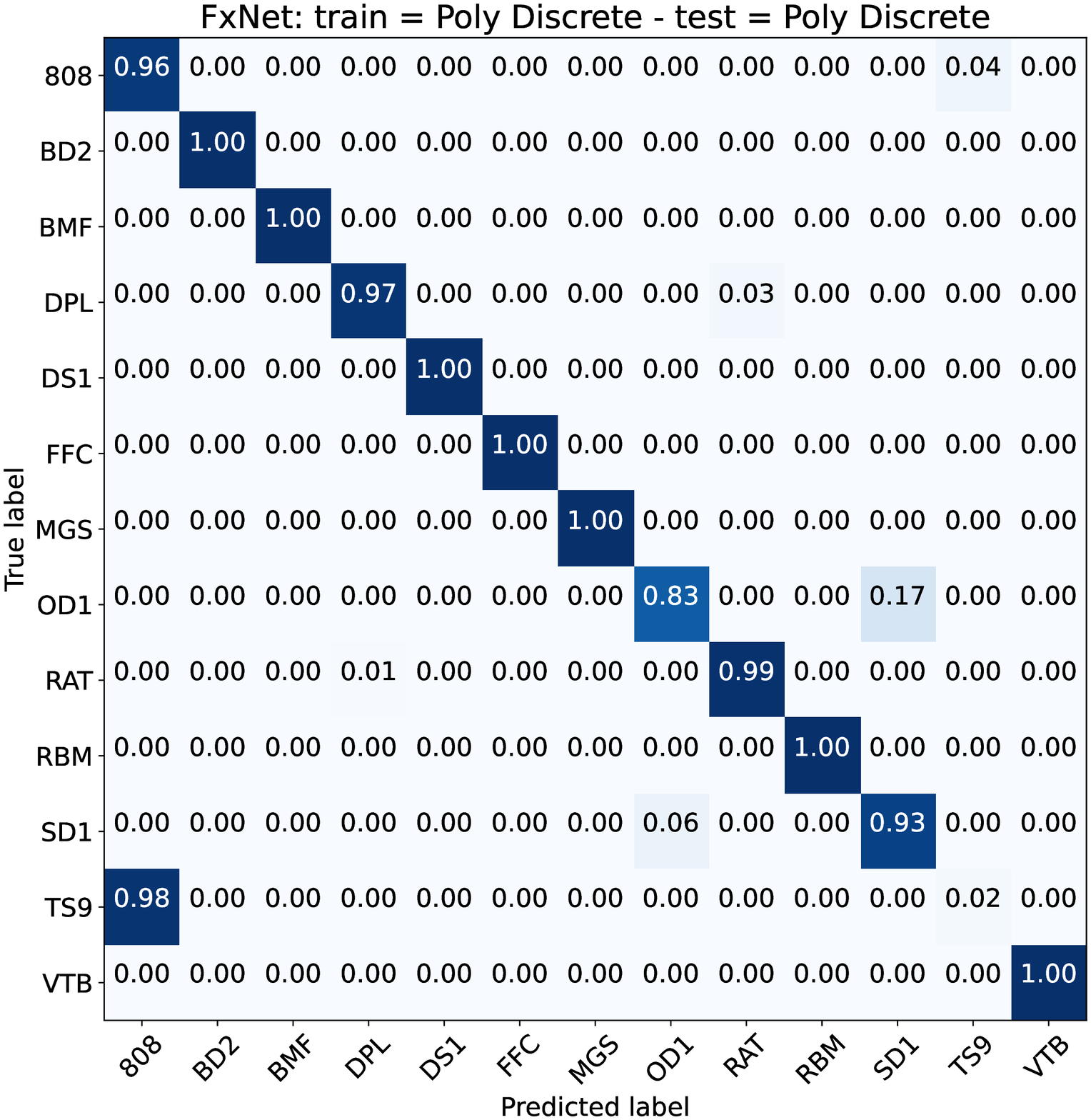}}
    \caption{confusion matrices for discrete datasets}
    \label{fig:fig1}
\end{figure*}

By analysing the confusion matrices, we can gain a better understanding about the networks' performance as well as the challenges behind the classification. Figure \ref{fig:fig1} shows the details for networks trained on discrete datasets. About 10\% of the errors in both datasets are due to the misclassification between 808 and TS9. The plugins are emulations of two overdrive effects from the same manufacturer (Ibanez TS808 and TS9). The two effects are supposed to have similar gain and frequency response, but, upon studying the circuits schematics, we noticed how there is actually no difference between the two.
Therefore - assuming the plugins are faithful models of the analogue circuits - it is plausible that our network would confuse samples from either effects. An explanation for the misclassification imbalance between monophonic and polyphonic samples might be related to the training procedure. We used batches of 100 samples, randomly selected across the whole dataset and without control for batch to dataset ratio over the 13 classes. Assuming identical plugins, the network will tend to classify samples from either as belonging to the class that was seen first or more during training.

Similar observations are valid for errors in classifying OD1 and SD1. Again, the plugins are emulations of effect pedals from the same manufacturer, and in their original analogue version use very similar designs. The two share the same clipping circuit; although, while the SD1 includes a tone control that combines a treble boosting first order shelving filter with a first order lowpass, the OD1 has no tone control and a fixed first order lowpass. Analysing the classification errors for the Mono Discrete dataset, we noticed how, of the 345 times the OD1 is classified as SD1: 31\% of the times is when the Gain control is set to 0.2, and another 31\% when Gain = 0.5. Ignoring the specific note being played, this result might be consequence of low gain settings, where the spectral differences might be too small. On the other hand, the SD1 is confused for the OD1 88 times, and all cases are from samples where the Tone is set to 0, 0.2 or 0.5. For low Tone values, the spectral differences might be unnoticeable, most of the high frequency harmonics might be filtered. In this case we did not observe correlation with the Gain control values.

% CONFUSION MATRICES

Another interesting example, more noticeable for the Mono Discrete dataset, is the misclassification of the DPL as RAT. In this case, we are referring to effects from different manufacturers which do share some circuit design choices. Although the DPL does not have a tone control, the two circuits use similar clipping stages and the same maximum gain. But, they differ in the filtering after the clipping section and in the type of clipping diodes: germanium in the DPL, silicon in the RAT; with the first type determining a softer clipping. Looking at the results we noticed how 72\% of the times the DPL is classified as RAT, the Gain control was set to 1.0. On the other hand, 60\% of the times the RAT was classified as DPL, the Gain was set to 0.5 and the Tone (a first order lowpass) to 0 (no high frequency attenuation). This might be related to the clipping diodes, a low gain with silicon diodes might be comparable to the softer clipping of germanium diodes.

It is also relevant to observe that 808, TS9, MGS and SD1, despite being based on the same circuit, with similar clipping sections and tone controls, are almost never confused.

\begin{figure*}
    \hspace{-0.4cm}
    \subfloat{\includegraphics[width=.515\linewidth]{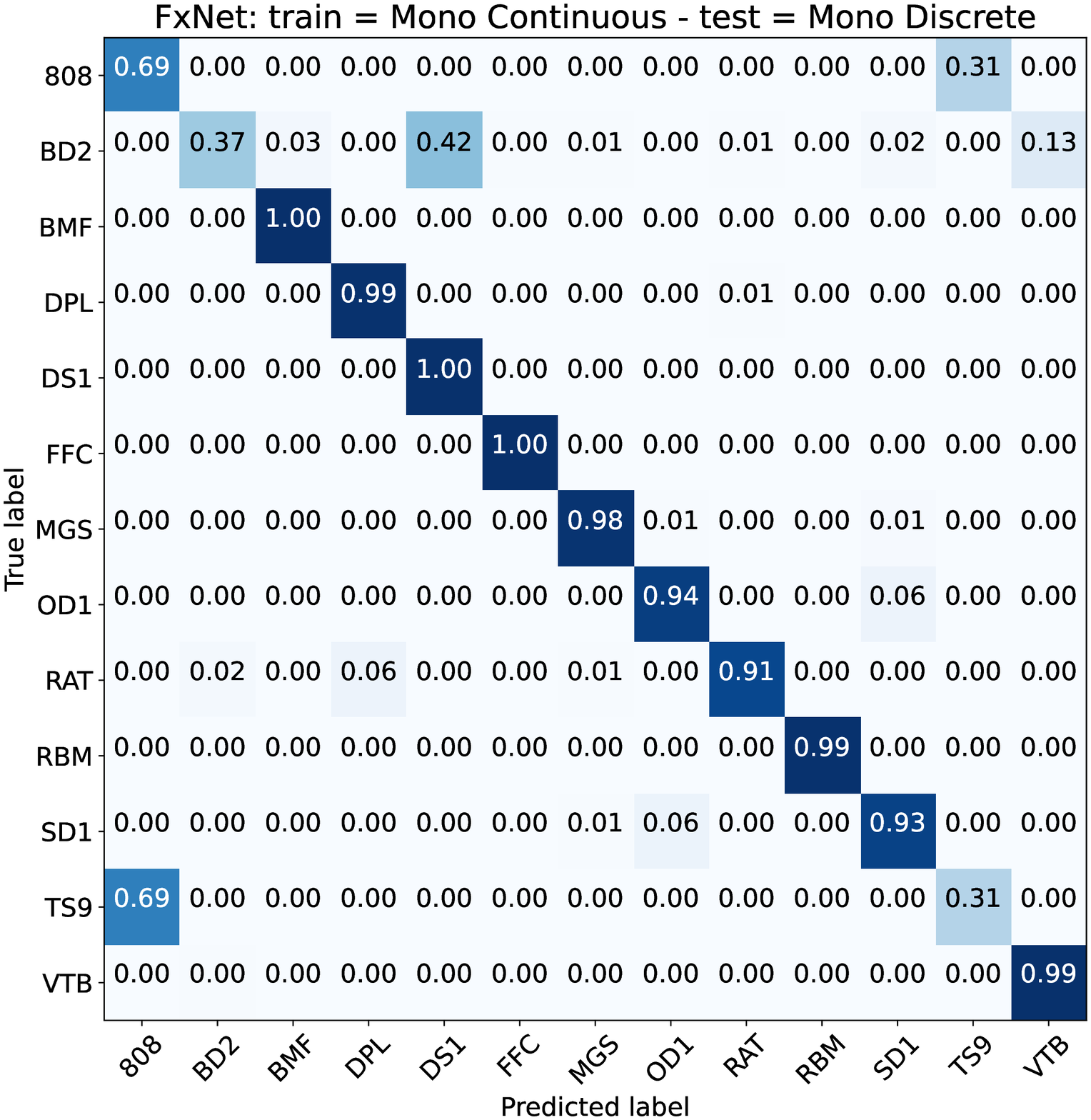}}
    \subfloat{\includegraphics[width=.515\linewidth]{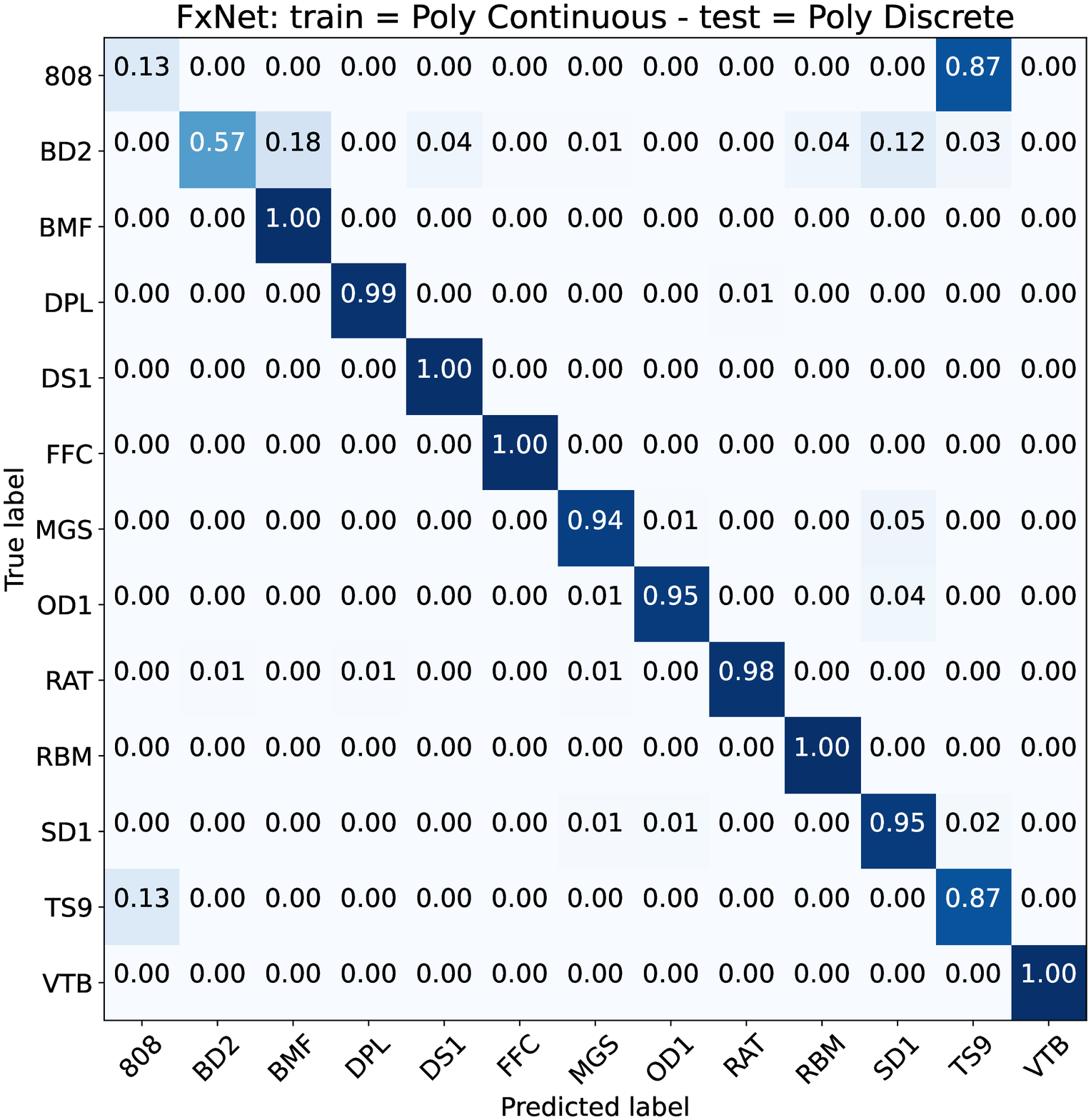}}
    \caption{confusion matrices for test on discrete datasets}
    \label{fig:fig2}
\end{figure*}

To analyse the performance of our classifiers trained on continuous datasets and tested on discrete ones, we refer to Figure \ref{fig:fig2}. It can be noticed how the errors are similar to the previous cases, although, a major impact on the performance is due to the misclassification of BD2. In this case, we could not identify correlations between the different circuit designs and/or the controls values.

% CONFUSION MATRICES

%===================================================%
%   RESULTS: SETTINGS ESTIMATION
%===================================================%
\subsection{Settings Estimation}
In this section we present the results for the settings estimation problem using the conditional network (SetNetCond) conditioned on the effect class ground truth. Table \ref{table:setnet_acc} shows the accuracy results, where all settings are estimated with an error below 0.1. Networks trained and tested on polyphonic datasets are the best performing, probably due to richer information content of the spectra with respect to monophonic samples. For both monophonic and polyphonic samples, the networks trained and tested on continuous settings reach higher accuracy than their counterparts trained and tested on discrete settings. This is somewhat surprising since the estimation of discrete values was expected to be a simpler problem to solve.

\begin{figure*}
    \hspace{-.3cm}
    \subfloat{\includegraphics[width=.255\linewidth]{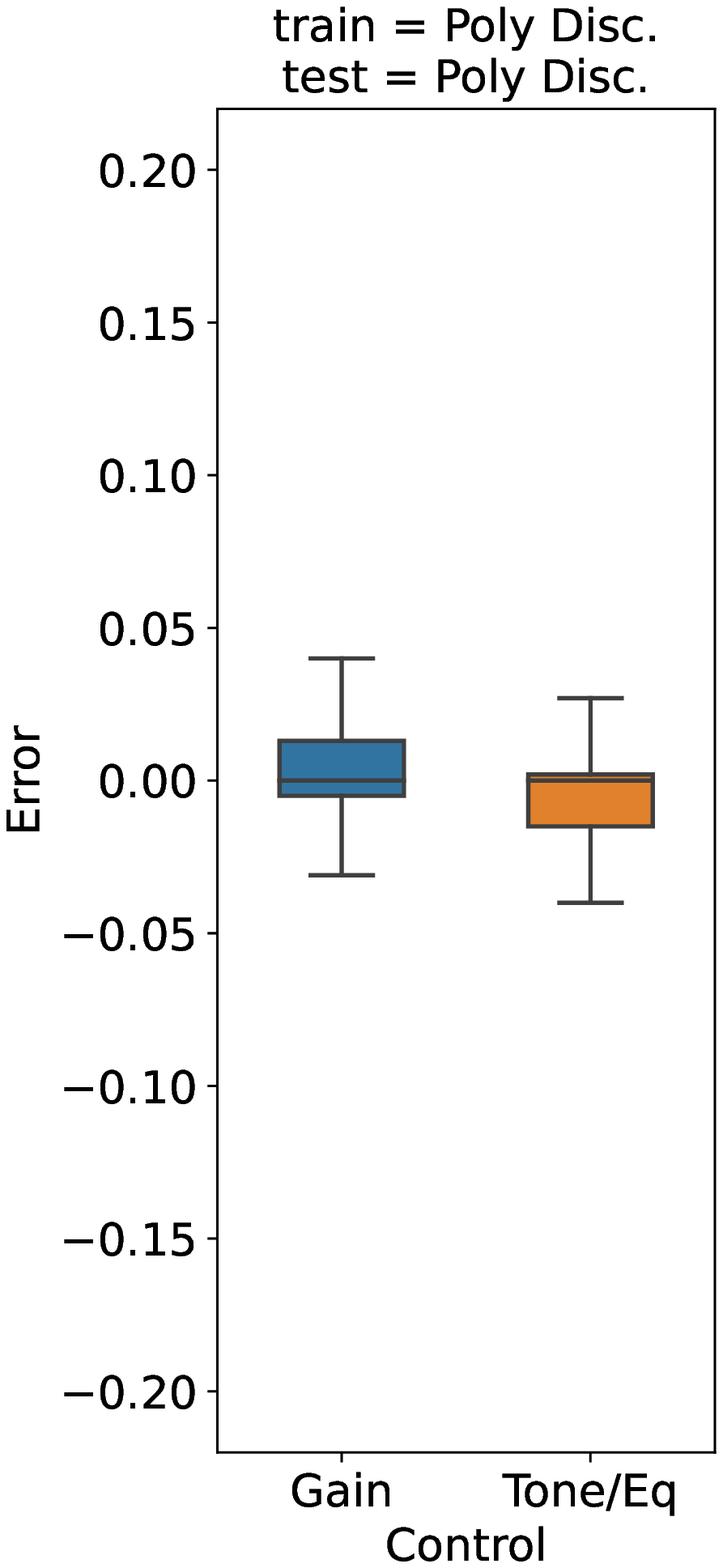}}
    \subfloat{\includegraphics[width=.255\linewidth]{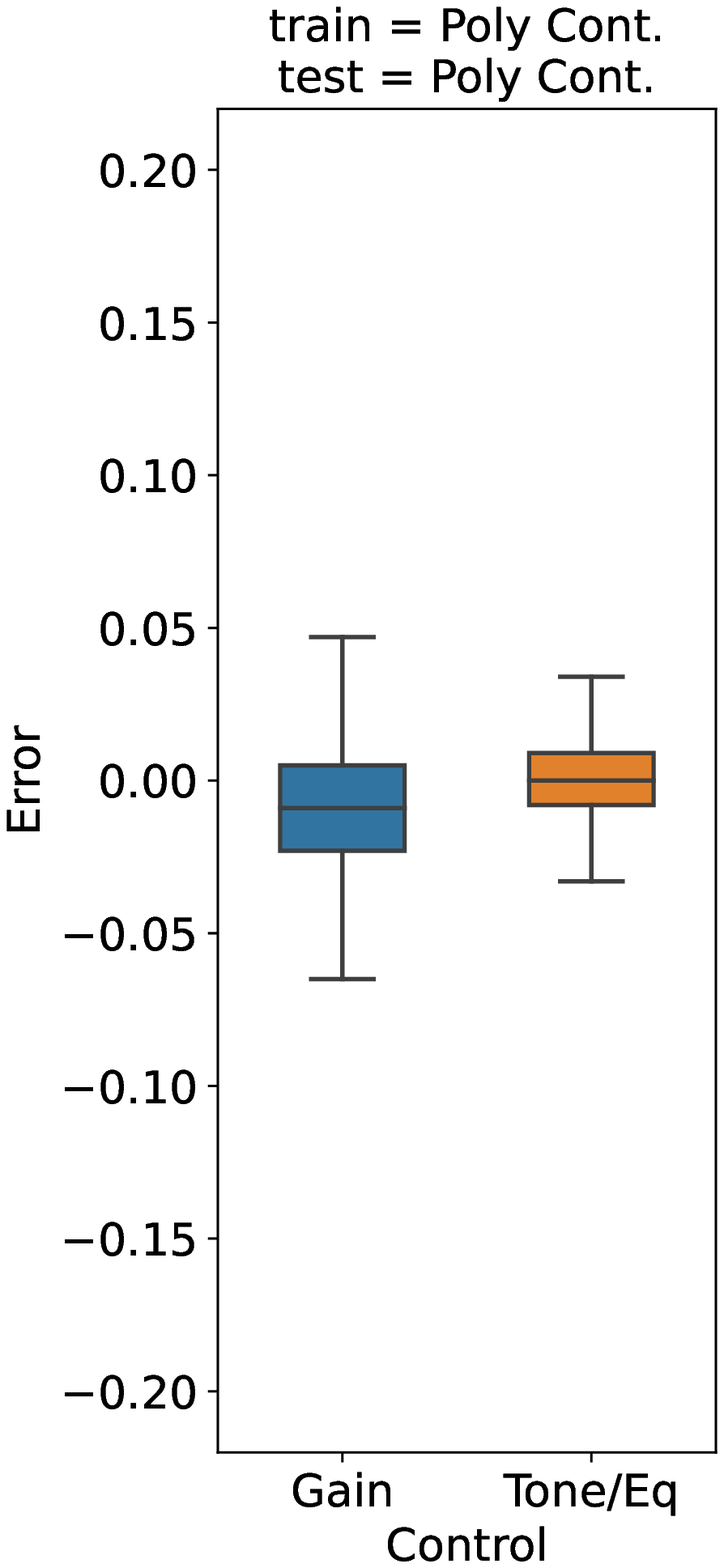}}
    \subfloat{\includegraphics[width=.255\linewidth]{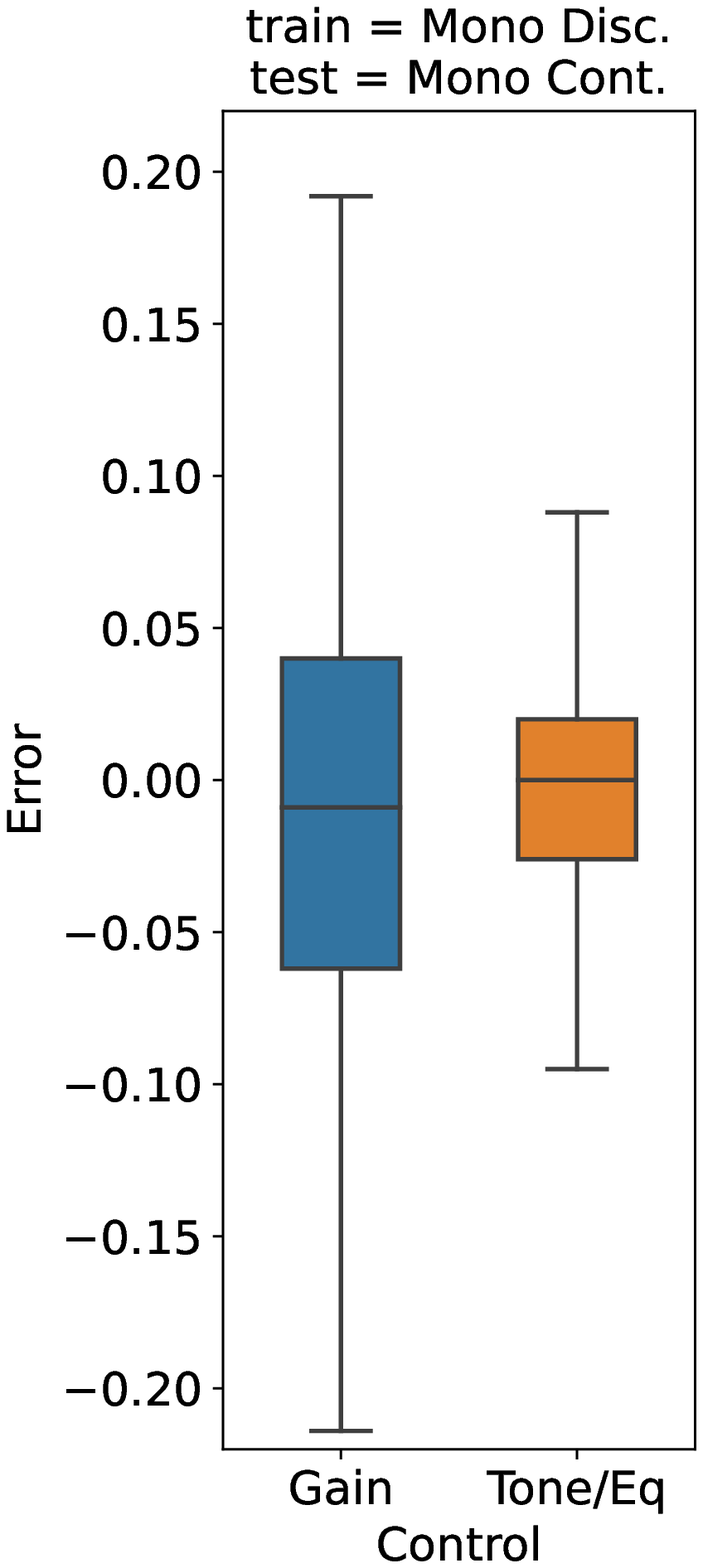}}
    \subfloat{\includegraphics[width=.255\linewidth]{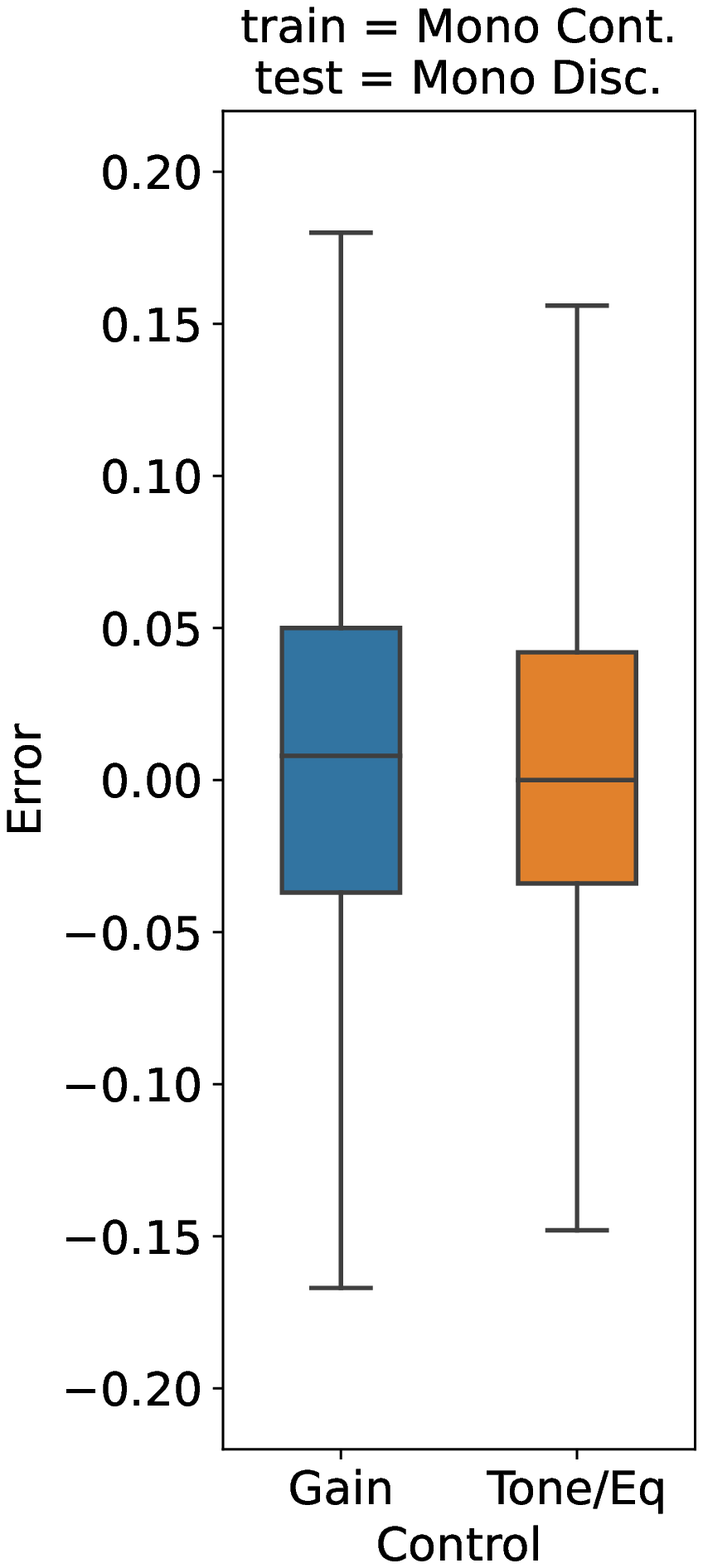}}
    \caption{settings estimation errors}
    \label{fig:fig3}
\end{figure*}

Further insights are offered by Table \ref{table:setnet_mae_rmse}, where we show mean absolute error (MAE) and root mean square error (RMSE) for the different training and testing configurations. In the majority of cases (12 out of 16) the MAE is below 0.05 and for all cases except one, the RMSE is below 0.1. We obtained the lowest errors when training and testing on polyphonic samples. The highest errors result from training on monophonic continuous and discrete samples and testing respectively on discrete and continuous ones. The table also includes the average errors for the Gain and Tone/Eq controls; the two are comparable, which shows that they present similar difficulty to the estimation. Figure \ref{fig:fig3} shows the box-plots for the best and worst cases highlighted in Table \ref{table:setnet_mae_rmse}.

\begin{figure*}[b!]
    \centering
    \subfloat{\includegraphics[width=.4\linewidth]{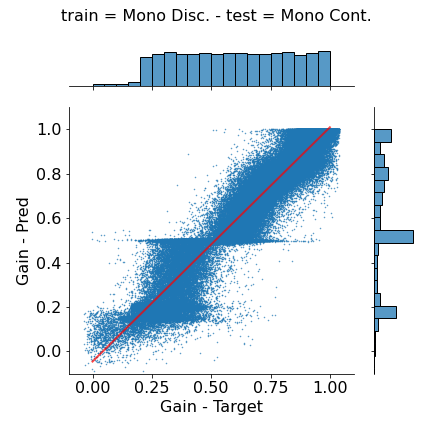}}
    \subfloat{\includegraphics[width=.4\linewidth]{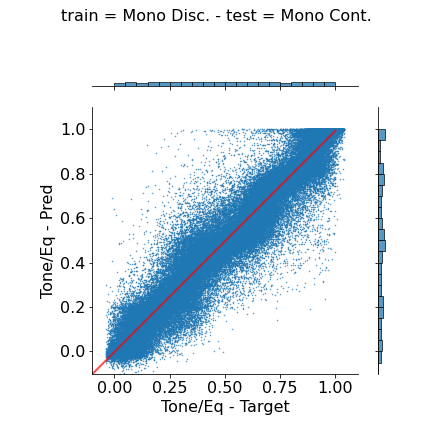}}
    \caption{scatter plots for settings estimation on continuous datasets}
    \label{fig:fig4}
% \end{figure*}
    
% \begin{figure*}
    \centering
    \subfloat{\includegraphics[width=.4\linewidth]{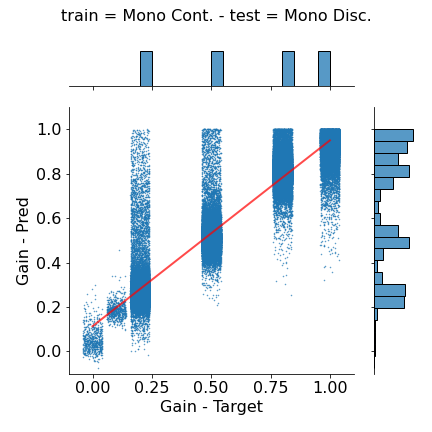}}
    \subfloat{\includegraphics[width=.4\linewidth]{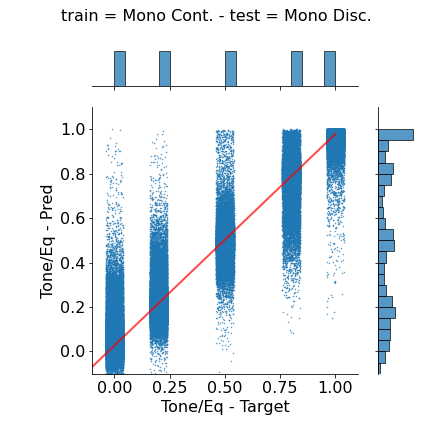}}
    \caption{scatter plots for settings estimation on discrete datasets}
    \label{fig:fig5}
\end{figure*}

Moreover, we wanted to analyse the generalisation capabilities of the networks trained on discrete settings, as well as the performance of networks trained on continuous settings on discrete ones. Figure \ref{fig:fig4} shows the scatter plots for the network trained on the Mono Discrete dataset when tested on the Mono Continuous dataset. For the Gain estimation we notice a bias towards the discrete values seen during training; also, the network manages to interpolate and estimate continuous values, but seems to do it satisfactorily only for Gain values above 0.5. The Tone estimation seems to be less affected, and the output approximates fairly well the input uniform distribution. An explanation for this difference might reside in the fact that for the Tone control we chose a balanced set of discrete values ([0.0, 0.2, 0.5, 0.8, 1.0]). For the Gain control this was not possible. As explained in section \ref{sec:dataset}, most distortion effects do not produce any perceivable timbral difference for low gain values and some introduce attenuation.

\begin{figure*}
    % \hspace{-0.3cm}
    \subfloat[]{\includegraphics[width=.26\linewidth]{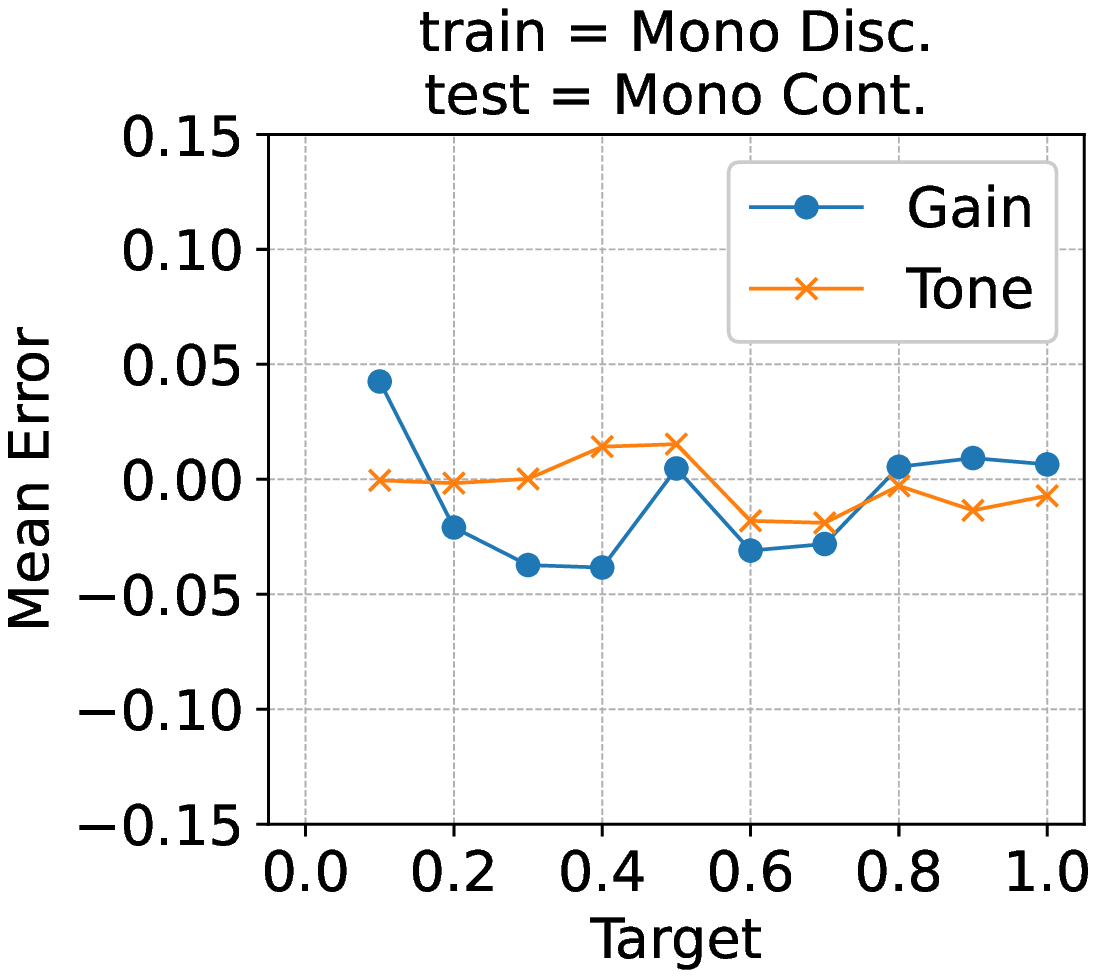}}
    % \hspace{-0.35cm}
    \subfloat[]{\includegraphics[width=.245\linewidth]{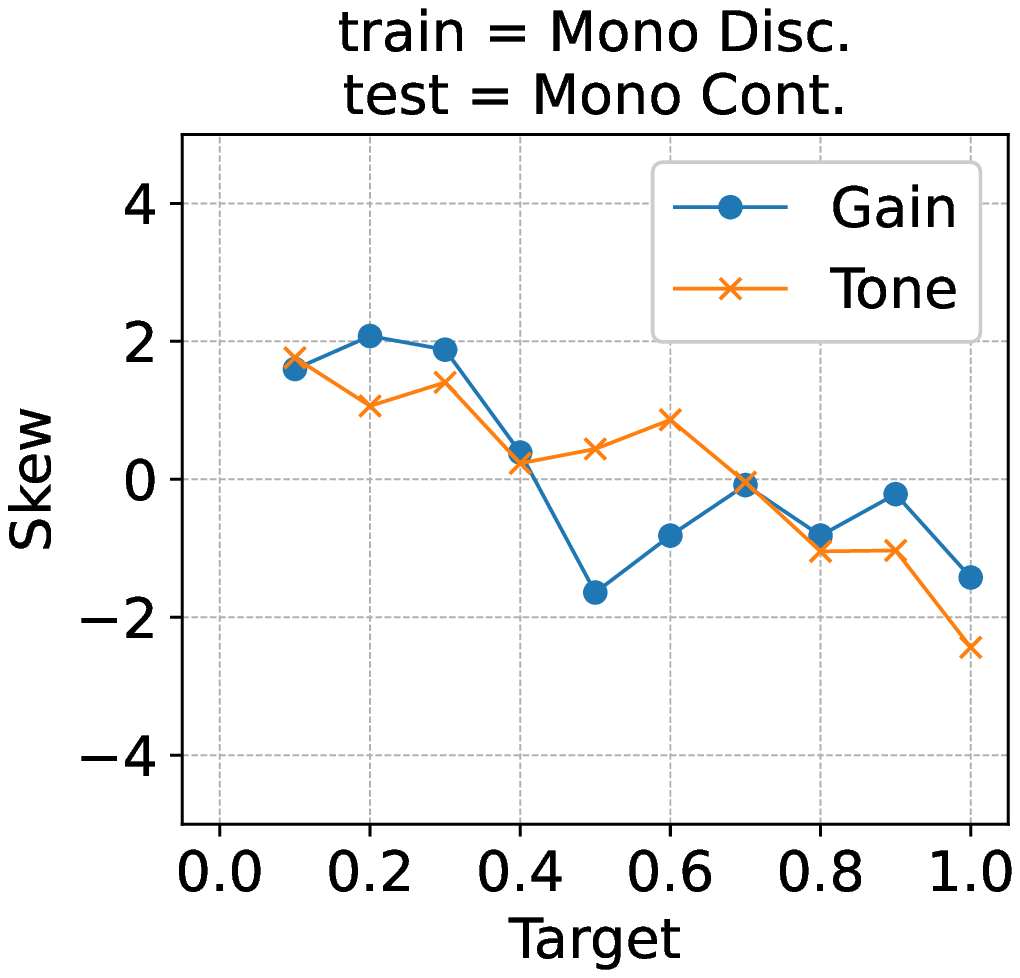}}
    % \hspace{-0.35cm}
    \subfloat[]{\includegraphics[width=.26\linewidth]{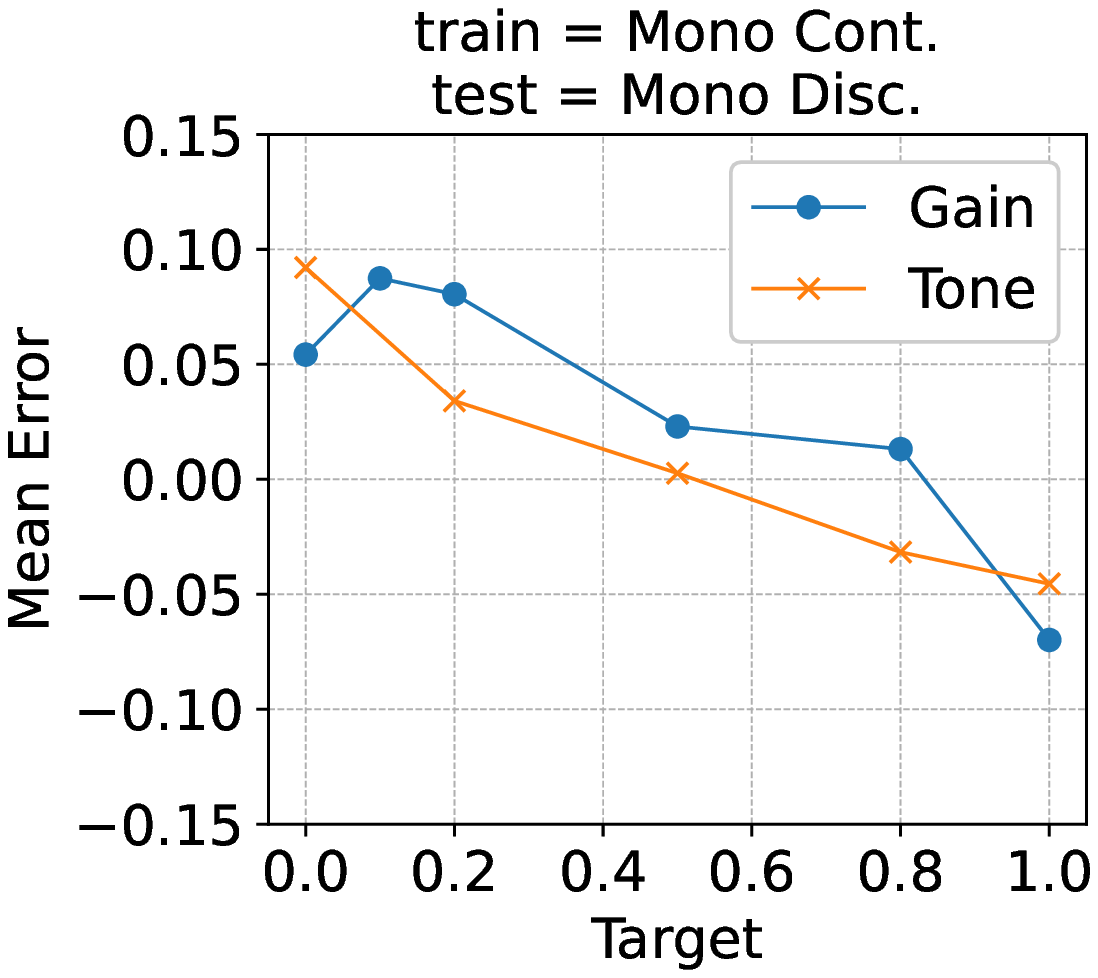}}
    % \hspace{-0.35cm}
    \subfloat[]{\includegraphics[width=.245\linewidth]{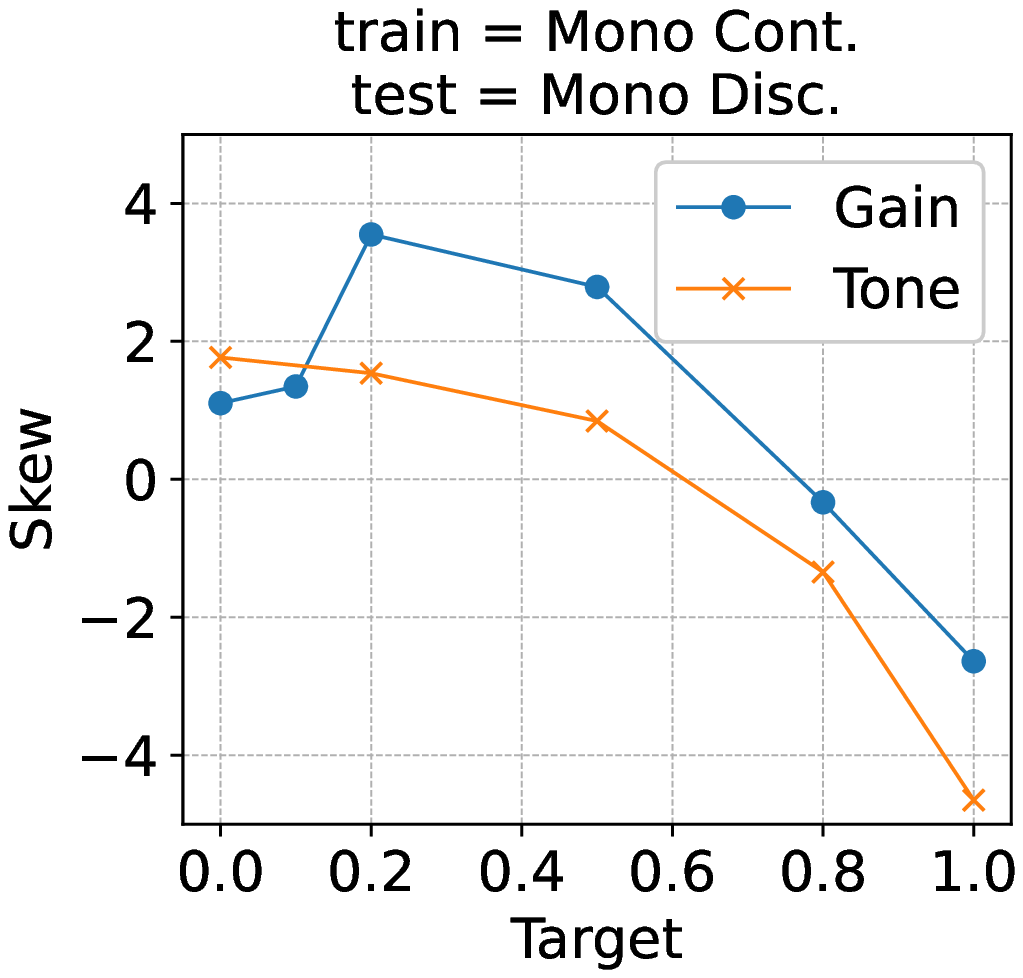}}
    \caption{mean error and skew for networks trained/tested on monophonic samples}
    \label{fig:fig6}
% \end{figure*}

% \begin{figure*}
    % \hspace{-0.3cm}
    \subfloat{\includegraphics[width=.26\linewidth]{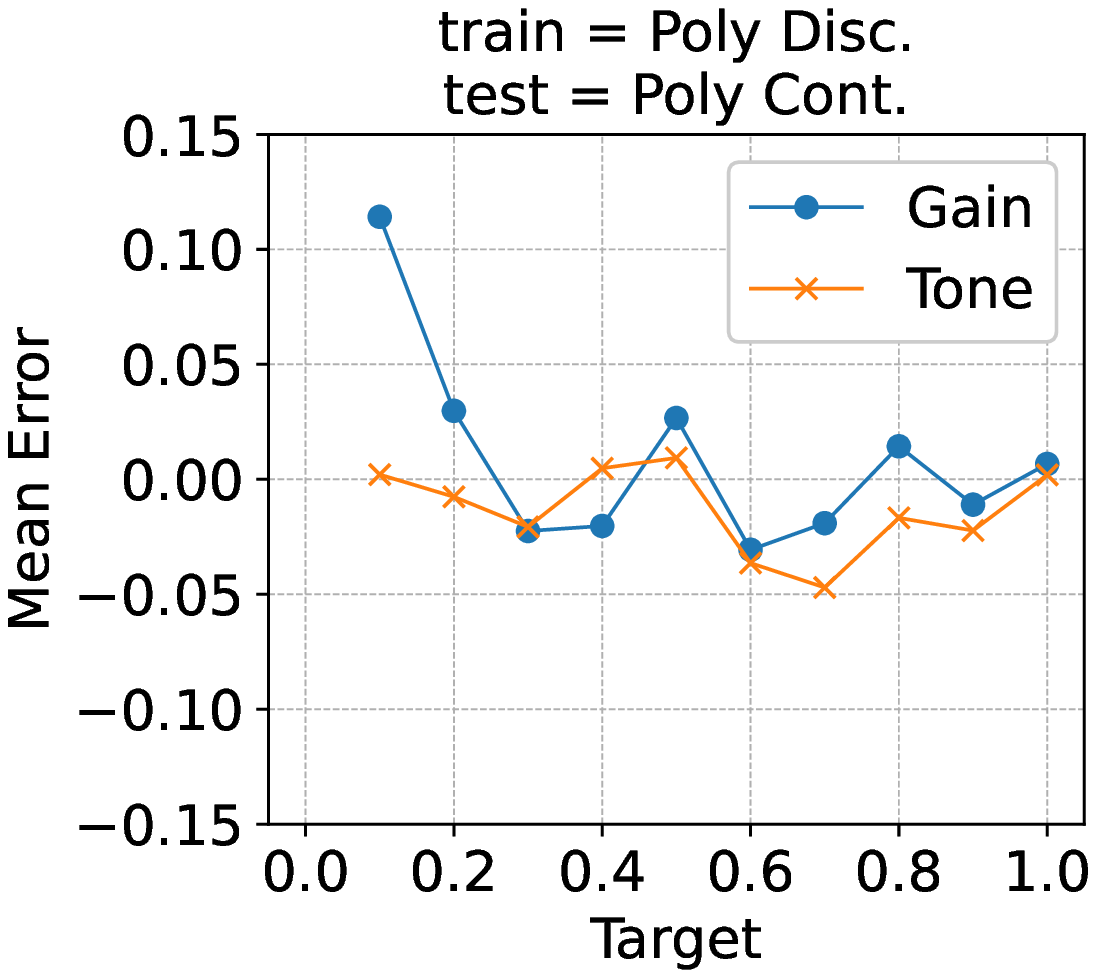}}
    % \hspace{-0.35cm}
    \subfloat{\includegraphics[width=.245\linewidth]{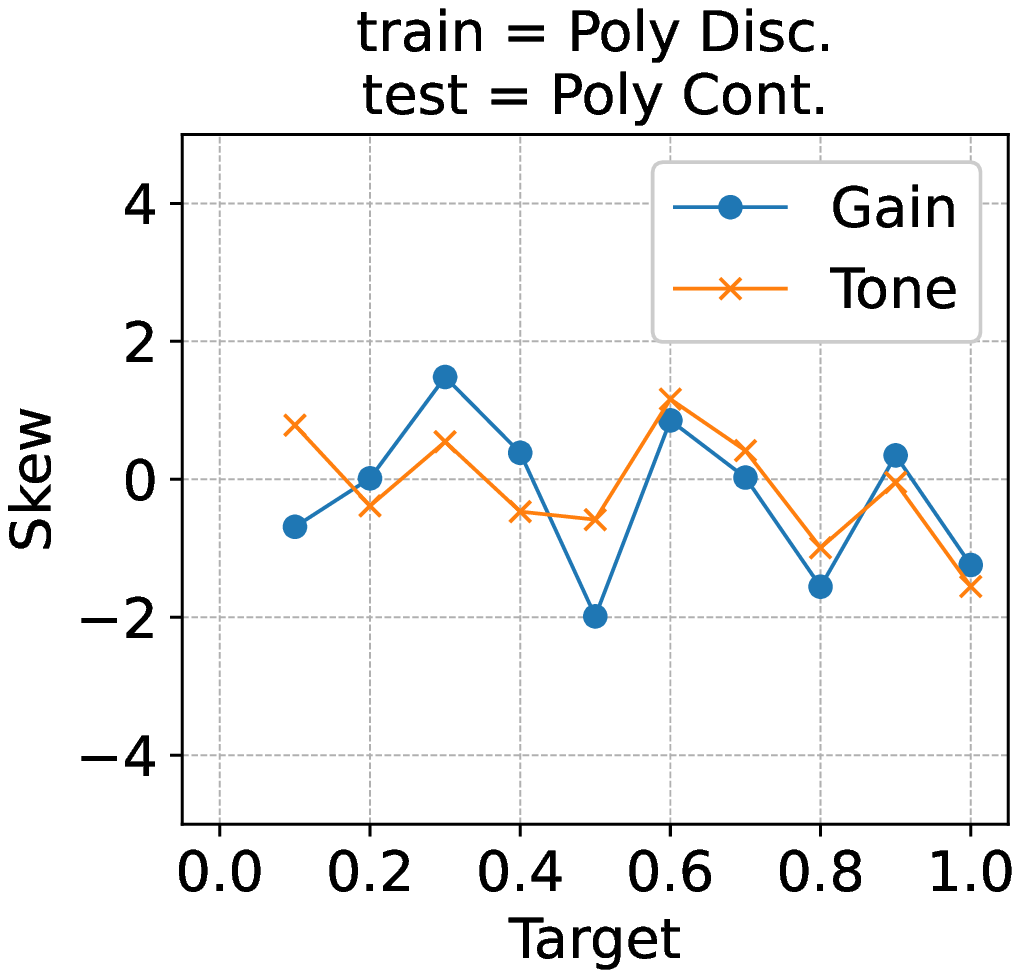}}
    % \hspace{-0.35cm}
    \subfloat{\includegraphics[width=.26\linewidth]{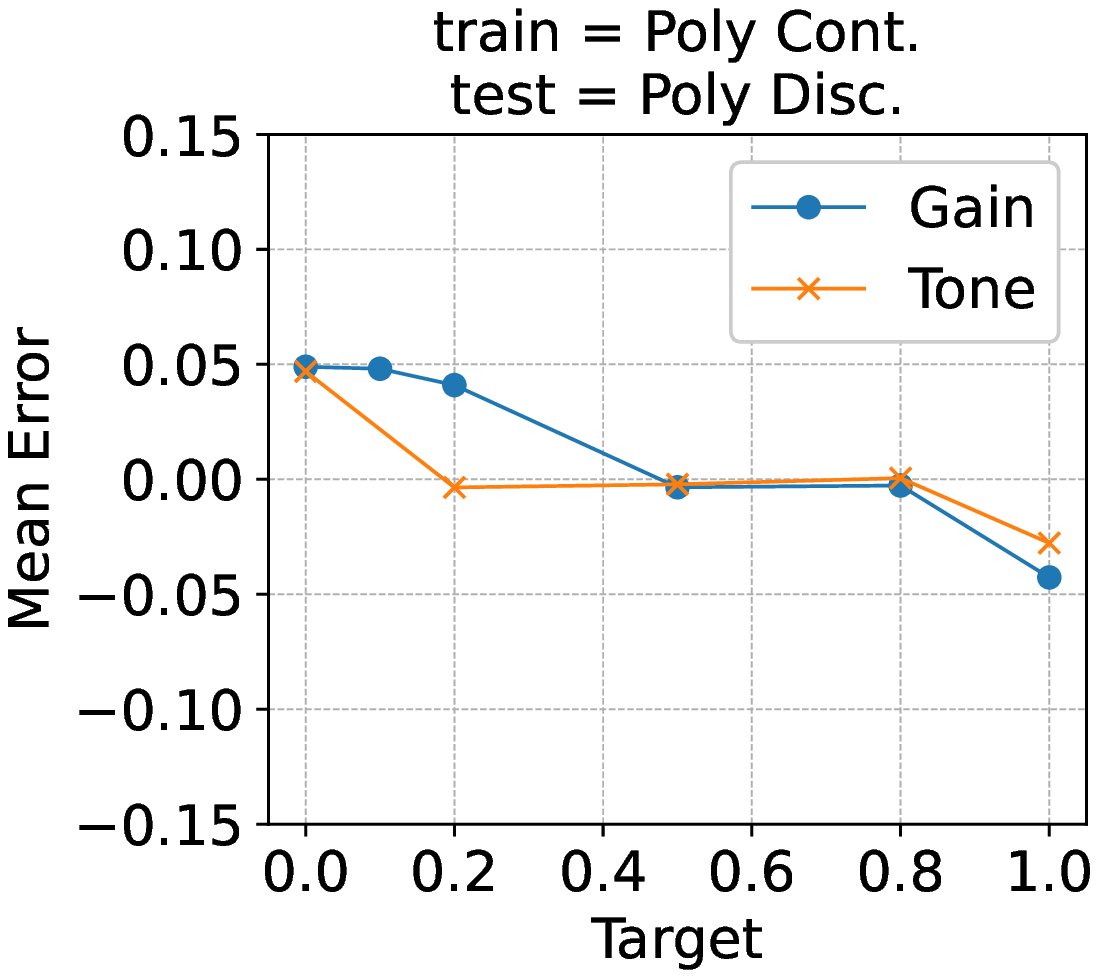}}
    % \hspace{-0.35cm}
    \subfloat{\includegraphics[width=.245\linewidth]{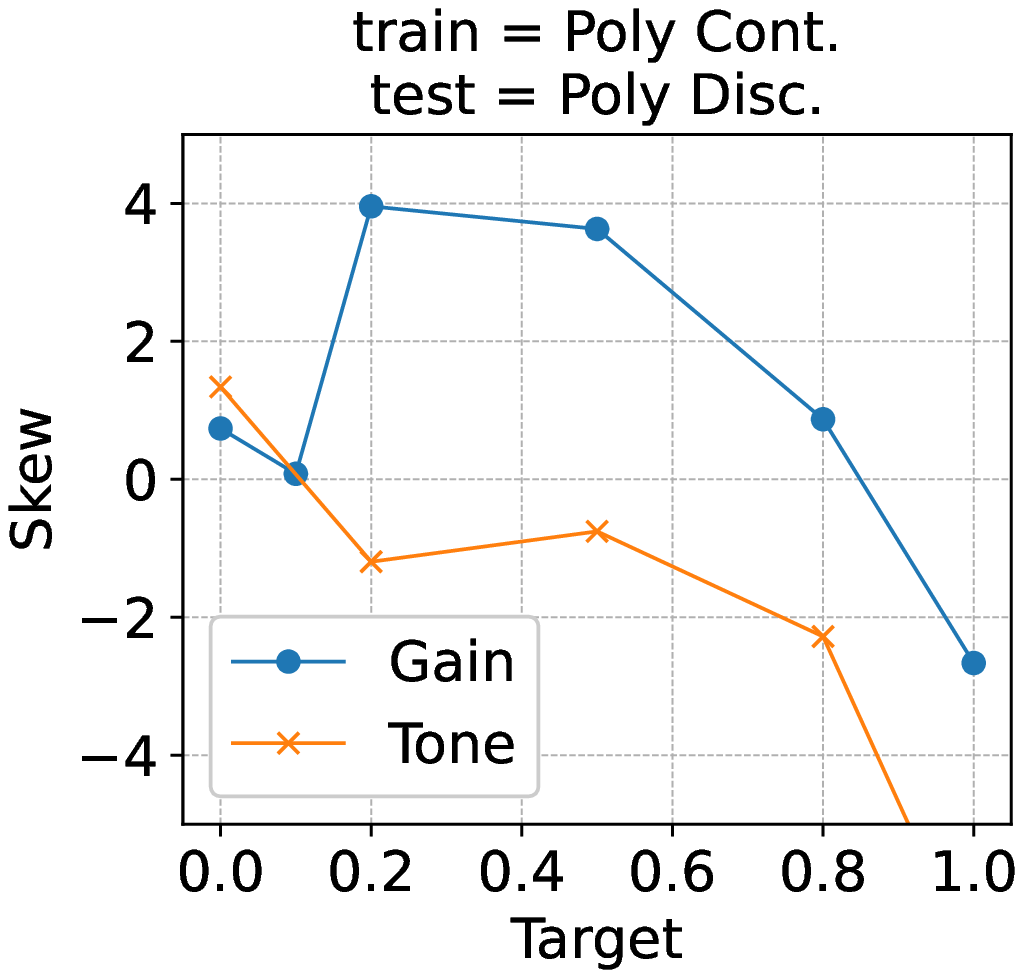}}
    \caption{mean error and skew for networks trained/tested on polyphonic samples}
    \label{fig:fig7}
\end{figure*}

The scatter plots for the network trained on the Mono Continuous and tested on the Mono Discrete datasets (Figure \ref{fig:fig5}) show some interesting behaviour. Tested on a discrete dataset, the network fails to maintain the performance at the same levels as on the continuous one. In particular, we notice a higher variance, especially for Gain = [0.2, 0.5] and for most of the Tone values. Also, a skew in the estimations' distributions is visible. To analyse these in more details, Figure \ref{fig:fig6} and \ref{fig:fig7} show the mean error and skew as a function of Gain and Tone for those networks trained on discrete datasets and tested on continuous ones and vice-versa. Except for the case of Gain values below 0.1 in Figure \ref{fig:fig7}a (train on Poly Discrete and test on Poly Continuous), all mean errors for tests on opposite datasets are below 0.1. With the same exception, the mean errors for training on discrete datasets and test on continuous ones are lower than 0.05 (Figures \ref{fig:fig6}a and \ref{fig:fig7}a). The skew for training on discrete and test on continuous datasets (Figures \ref{fig:fig6}b and \ref{fig:fig7}b) is in many cases lower than the inverse conditions (Figures \ref{fig:fig6}b, \ref{fig:fig6}d and \ref{fig:fig7}b, \ref{fig:fig7}d).

To conclude, even if the networks perform better on continuous datasets, there seems to be an argument for considering discrete values. A dataset which uses discrete values for the independent variables is easier to design, control, analyse and to extend or reduce. In our specific application, an estimation error within 0.1 of the target value or some bias is acceptable.
Also, the use of balanced values or a form of regularisation in the cost function could help the interpolation in case of estimation on continuous values or unseen data. 

\begingroup
\setlength{\tabcolsep}{4pt} % Default value: 6pt
\renewcommand{\arraystretch}{1} % Default value: 1
\begin{table}
    \centering
    \small
    \begin{tabular}{|c|c|c|c|c|}
    \hline
        \backslashbox[]{Train}{Test}
            & \textbf{Mono Disc.} 
                & \textbf{Mono Cont.} 
                    & \textbf{Poly Disc.} 
                        & \textbf{Poly Cont.}\\
        \hline
        \textbf{Mono Dis.}
            &  80.06
                & 68.56
                    & ---
                        & ---\\
        \textbf{Mono Cont.}
            &  68.51
                & 85.14
                    & ---
                        & ---\\
        \textbf{Poly Disc.}
            & ---
                & ---
                    & 90.75
                        & 75.74\\
        \textbf{Poly Cont.}
            & ---
                & ---
                    & 88.93
                        & 97.01\\
        
        \hline
    \end{tabular}
    \caption{SetNetCond accuracy (\%)}
    \label{table:setnet_acc}
\end{table}
\endgroup
\begin{table}
\small
\centering
\setlength{\tabcolsep}{4pt}
\begin{tabular}{|c|c|c|c|}
\hline
    \multirow{2}{*}{\textbf{Train Set}} 
            & \textbf{Gain} 
                & \textbf{Tone/Eq} 
                    & \multirow{2}{*}{\textbf{Test Set}} \\
            & MAE \textit{(RMSE)}
                & MAE \textit{(RMSE)}
                    & \\
\hline
    \multirow{2}{*}{Mono Disc.} 
        & 0.030 \textit{(0.061)}
            & 0.039 \textit{(0.070)} 
                & Mono Disc.\\
        & \textbf{0.064 \textit{(0.084)}}
            & \textbf{0.044 \textit{(0.080)}}
                & Mono Cont.\\
\hline
    \multirow{2}{*}{Mono Cont.}
        & \textbf{0.062 \textit{(0.096)}}  
            & \textbf{0.067 \textit{(0.108)}}
                & Mono Disc.\\
        & 0.033 \textit{(0.045)}
            & 0.039 \textit{(0.072)} 
                & Mono Cont.\\
\hline
    \multirow{2}{*}{Poly Disc.} 
        & \textbf{0.017 \textit{(0.033)}}  
            & \textbf{0.024 \textit{(0.047)}}  
                &   Poly Cont.\\
     
        & 0.055 \textit{(0.070)}
            & 0.038 \textit{(0.062)} 
                &   Poly Cont.\\
\hline
    \multirow{2}{*}{Poly Cont.} 
        & 0.036 \textit{(0.063)}  
            & 0.036 \textit{(0.062)}  
                & Poly Disc.\\
     
        & \textbf{0.020 \textit{(0.028)}}  
            & \textbf{0.019 \textit{(0.034)}}  
                & Poly Cont.\\
\hline
    \textbf{Avg}
        & 0.040 \textit{(0.060)}
            & 0.038 \textit{(0.066)}
                & \\
\hline
\end{tabular}
\caption{SetNetCond errors}
\label{table:setnet_mae_rmse}
\end{table}

% BOX PLOTS

% SCATTER PLOTS

% MEAN ERROR and SKEW PLOTS

%===================================================%
%   CONCLUSIONS, FUTURE WORK
%===================================================%
\section{Conclusions and Future Work}
\label{sec:conclusion}
In this paper we introduced a CNN-based model for the classification of guitar effects and estimation of their parameters. Using electric guitar recordings of single notes, 2-notes intervals and 3- or 4- notes chords; we were able to classify with high accuracy samples processed with 13 overdrive, distortion and fuzz plugins. The plugins we used are emulations of some of the most famous and commonly used analogue guitar effect pedals. We were also able to estimate the parameters settings with high precision. For this work, we generated a dataset of processed monophonic and polyphonic samples; the plugins' settings were either combinations of discrete values commonly used by musicians or continuous values randomly extracted from a uniform distribution. This allowed us to gain further understanding about the performance on datasets generated using discrete or continuous independent variables. We described the benefits of discrete datasets and postulated about the possibility of obtaining equally high performances.

To the best of our knowledge, our work is the first attempt at both classification and parameter estimation of specific guitar effect units. Being able to extrapolate information about the specific signal chain used in a recording could benefit music production; artists, engineers and producers often rely on "reference" sounds from other recordings. This knowledge could also be applied to automatic music transcription \cite{benetos2013automatic}, music education \cite{dittmar2012music} or musicology \cite{abesser2016score}. Musicians, genres and styles are identified by and associated with specific sounds and effects; this could be applied to intelligent music search and recommendation systems.

Future work might branch out in different directions: our model could be compared with human performance; the architecture and datasets could be extended for higher accuracy and better generalisation on unseen data; and the research could be extended to other nonlinear units, different categories of effects, or tested on other guitars or instruments.

%===================================================%
%   REPOS
%===================================================%
\section{Links}
\label{sec:repos}
Code:
\begin{itemize}
    \item \href{https://github.com/mcomunita/gfx\_classifier}{https://github.com/mcomunita/gfx\_classifier}
    \item \href{https://github.com/mcomunita/gfx_classifier\_models\_and_results}{https://github.com/mcomunita/gfx\_classifier\_models\_and\_results}
\end{itemize}
Dataset:
\begin{itemize}
    \item Mono Cont. - \href{https://doi.org/10.5281/zenodo.4296040}{https://doi.org/10.5281/zenodo.4296040}
    \item Mono Disc. - \href{https://doi.org/10.5281/zenodo.4298000}{https://doi.org/10.5281/zenodo.4298000}
    \item Poly Cont. - \href{https://doi.org/10.5281/zenodo.4298017}{https://doi.org/10.5281/zenodo.4298017}
    \item Poly Disc. - \href{https://doi.org/10.5281/zenodo.4298025}{https://doi.org/10.5281/zenodo.4298025}
\end{itemize}

%===================================================%
%   ACKNOWLEDGEMENT
%===================================================%
\section{Acknowledgement}
\label{sec:acknowledgement}
This study has been funded by UKRI and EPSRC as part of the “UKRI Centre for Doctoral Training in Artificial Intelligence and Music”, under grant EP/S022694/1.

%===================================================%
%   REFERENCES
%===================================================%
\bibliographystyle{unsrt}  
\bibliography{article}

\begin{thebibliography}{10}

\bibitem{case2010recording}
Alex Case.
\newblock Recording electric guitar—the science and the myth.
\newblock {\em Journal of the Audio Engineering Society}, 58(1/2):80--83, 2010.

\bibitem{blair2015southern}
John Blair.
\newblock {\em Southern California surf music, 1960-1966}.
\newblock Arcadia Publishing, 2015.

\bibitem{williams2012tubby}
Sean Williams.
\newblock Tubby's dub style: The live art of record production.
\newblock 2012.

\bibitem{prown2003gear}
Pete Prown and Lisa Sharken.
\newblock {\em Gear secrets of the guitar legends: how to sound like your
  favorite players}.
\newblock Hal Leonard Corporation, 2003.

\bibitem{zolzer2011dafx}
Udo Z{\"o}lzer.
\newblock {\em DAFX: digital audio effects}.
\newblock John Wiley \& Sons, 2011.

\bibitem{pakarinen2009review}
Jyri Pakarinen and David~T Yeh.
\newblock A review of digital techniques for modeling vacuum-tube guitar
  amplifiers.
\newblock {\em Computer Music Journal}, 33(2):85--100, 2009.

\bibitem{herrera2003automatic}
Perfecto Herrera-Boyer, Geoffroy Peeters, and Shlomo Dubnov.
\newblock Automatic classification of musical instrument sounds.
\newblock {\em Journal of New Music Research}, 32(1):3--21, 2003.

\bibitem{benetos2013automatic}
Emmanouil Benetos, Simon Dixon, Dimitrios Giannoulis, Holger Kirchhoff, and
  Anssi Klapuri.
\newblock Automatic music transcription: challenges and future directions.
\newblock {\em Journal of Intelligent Information Systems}, 41(3):407--434,
  2013.

\bibitem{kehling2014automatic}
Christian Kehling, Jakob Abe{\ss}er, Christian Dittmar, and Gerald Schuller.
\newblock Automatic tablature transcription of electric guitar recordings by
  estimation of score-and instrument-related parameters.
\newblock In {\em DAFx}, pages 219--226, 2014.

\bibitem{dittmar2012music}
Christian Dittmar, Estefan{\'\i}a Cano, Jakob Abe{\ss}er, and Sascha
  Grollmisch.
\newblock Music information retrieval meets music education.
\newblock In {\em Dagstuhl Follow-Ups}, volume~3. Schloss
  Dagstuhl-Leibniz-Zentrum fuer Informatik, 2012.

\bibitem{abesser2016score}
Jakob Abe{\ss}er, Klaus Frieler, Estefan{\'\i}a Cano, Martin Pfleiderer, and
  Wolf-Georg Zaddach.
\newblock Score-informed analysis of tuning, intonation, pitch modulation, and
  dynamics in jazz solos.
\newblock {\em IEEE/ACM Transactions on Audio, Speech, and Language
  Processing}, 25(1):168--177, 2016.

\bibitem{barbancho2009pitch}
Isabel Barbancho, Lorenzo~J Tardon, Ana~M Barbancho, and Simone Sammartino.
\newblock Pitch and played string estimation in classic and acoustic guitars.
\newblock In {\em Audio Engineering Society Convention 126}, 2009.

\bibitem{abesser2012automatic}
Jakob Abe{\ss}er.
\newblock Automatic string detection for bass guitar and electric guitar.
\newblock In {\em International Symposium on Computer Music Modeling and
  Retrieval}, pages 333--352. Springer, 2012.

\bibitem{dittmar2013real}
Christian Dittmar, Andreas M{\"a}nnchen, and Jakob Abeber.
\newblock Real-time guitar string detection for music education software.
\newblock In {\em 2013 14th International Workshop on Image Analysis for
  Multimedia Interactive Services (WIAMIS)}, pages 1--4. IEEE, 2013.

\bibitem{geib2017automatic}
Tobias Geib, Maximilian Schmitt, and Bj{\"o}rn Schuller.
\newblock Automatic guitar string detection by string-inverse frequency
  estimation.
\newblock {\em INFORMATIK 2017}, 2017.

\bibitem{hjerrild2019estimation}
Jacob~M{\o}ller Hjerrild and Mads~Gr{\ae}sb{\o}ll Christensen.
\newblock Estimation of guitar string, fret and plucking position using
  parametric pitch estimation.
\newblock In {\em ICASSP 2019-2019 IEEE International Conference on Acoustics,
  Speech and Signal Processing (ICASSP)}, pages 151--155. IEEE, 2019.

\bibitem{hjerrild2019physical}
Jacob~M{\o}ller Hjerrild, Silvin Willemsen, and Mads~Gr{\ae}sb{\o}ll
  Christensen.
\newblock Physical models for fast estimation of guitar string, fret and
  plucking position.
\newblock In {\em 2019 IEEE Workshop on Applications of Signal Processing to
  Audio and Acoustics (WASPAA)}, pages 155--159. IEEE, 2019.

\bibitem{mohamad2017pickup2}
Zulfadhli Mohamad, Simon Dixon, and Christopher Harte.
\newblock Pickup position and plucking point estimation on an electric guitar
  via autocorrelation.
\newblock {\em The Journal of the Acoustical Society of America},
  142(6):3530--3540, 2017.

\bibitem{mohamad2017pickup}
Zulfadhli Mohamad, Simon Dixon, and Christopher Harte.
\newblock Pickup position and plucking point estimation on an electric guitar.
\newblock In {\em 2017 IEEE International Conference on Acoustics, Speech and
  Signal Processing (ICASSP)}, pages 651--655. IEEE, 2017.

\bibitem{abesser2010feature}
Jakob Abe{\ss}er, Hanna Lukashevich, and Gerald Schuller.
\newblock Feature-based extraction of plucking and expression styles of the
  electric bass guitar.
\newblock In {\em 2010 IEEE International Conference on Acoustics, Speech and
  Signal Processing}, pages 2290--2293. IEEE, 2010.

\bibitem{schuller2015parameter}
Gerald Schuller, Jakob Abe{\ss}er, and Christian Kehling.
\newblock Parameter extraction for bass guitar sound models including playing
  styles.
\newblock In {\em 2015 IEEE International Conference on Acoustics, Speech and
  Signal Processing (ICASSP)}, pages 404--408. IEEE, 2015.

\bibitem{su2014sparse}
Li~Su, Li-Fan Yu, and Yi-Hsuan Yang.
\newblock Sparse cepstral, phase codes for guitar playing technique
  classification.
\newblock In {\em ISMIR}, pages 9--14, 2014.

\bibitem{wang2020spectral}
Chien-Yao Wang, Pao-Chi Chang, Jian-Jiun Ding, Tzu-Chiang Tai, Andri Santoso,
  Yu-Ting Liu, and Jia-Ching Wang.
\newblock Spectral-temporal receptive field-based descriptors and hierarchical
  cascade deep belief network for guitar playing technique classification.
\newblock {\em IEEE Transactions on Cybernetics}, 2020.

\bibitem{foulon2013automatic}
Raphael Foulon, Pierre Roy, and Fran{\c{c}}ois Pachet.
\newblock Automatic classification of guitar playing modes.
\newblock In {\em International Symposium on Computer Music Multidisciplinary
  Research}, pages 58--71. Springer, 2013.

\bibitem{barbancho2011automatic}
Ana~M Barbancho, Anssi Klapuri, Lorenzo~J Tard{\'o}n, and Isabel Barbancho.
\newblock Automatic transcription of guitar chords and fingering from audio.
\newblock {\em IEEE Transactions on Audio, Speech, and Language Processing},
  20(3):915--921, 2011.

\bibitem{dosenbach2008identification}
Kerstin Dosenbach, Wolfgang Fohl, and Andreas Meisel.
\newblock Identification of individual guitar sounds by support vector
  machines.
\newblock In {\em Proceedings of the Conference on Digital Audio Effects
  (DAFx)}, 2008.

\bibitem{johnson2015guitar}
David Johnson and George Tzanetakis.
\newblock Guitar model recognition from single instrument audio recordings.
\newblock In {\em 2015 IEEE Pacific Rim Conference on Communications, Computers
  and Signal Processing (PACRIM)}, pages 370--375. IEEE, 2015.

\bibitem{profeta2019feature}
Renato de Castro~Rabelo Profeta and Gerald Schuller.
\newblock Feature-based classification of electric guitar types.
\newblock In {\em Joint European Conference on Machine Learning and Knowledge
  Discovery in Databases}, pages 478--484. Springer, 2019.

\bibitem{profeta2019comparison}
Renato Profeta and Gerald Schuller.
\newblock Comparison of human and machine recognition of electric guitar types.
\newblock In {\em Audio Engineering Society Convention 147}. Audio Engineering
  Society, 2019.

\bibitem{stein2010automatic}
Michael Stein, Jakob Abe{\ss}er, Christian Dittmar, and Gerald Schuller.
\newblock Automatic detection of audio effects in guitar and bass recordings.
\newblock In {\em Audio Engineering Society Convention 128}, 2010.

\bibitem{stein2010automatic2}
Michael Stein.
\newblock Automatic detection of multiple, cascaded audio effects in guitar
  recordings.
\newblock In {\em Proceedings of the 13th International Conference on Digital
  Audio Effects (DAFx)}, pages 4--7, 2010.

\bibitem{eichas2015feature}
Felix Eichas, Marco Fink, and Udo Z{\"o}lzer.
\newblock Feature design for the classification of audio effect units by
  input/output measurements.
\newblock {\em DAFx-15}, 2015.

\bibitem{schmitt2017recognising}
Maximilian Schmitt and Bj{\"o}rn Schuller.
\newblock Recognising guitar effects-which acoustic features really matter?
\newblock {\em INFORMATIK 2017}, 2017.

\bibitem{eichas2018gray}
Felix Eichas and Udo Z{\"o}lzer.
\newblock Gray-box modeling of guitar amplifiers.
\newblock {\em Journal of the Audio Engineering Society}, 66(12):1006--1015,
  2018.

\bibitem{jurgens2020recognizing}
Henrik J{\"u}rgens, Reemt Hinrichs, and J{\"o}rn Ostermann.
\newblock Recognizing guitar effects and their parameter settings.

\bibitem{mcfee2015librosa}
Brian McFee, Colin Raffel, Dawen Liang, Daniel~PW Ellis, Matt McVicar, Eric
  Battenberg, and Oriol Nieto.
\newblock librosa: Audio and music signal analysis in python.
\newblock In {\em Proceedings of the 14th python in science conference},
  volume~8, pages 18--25, 2015.

\bibitem{abesser2020review}
Jakob Abe{\ss}er.
\newblock A review of deep learning based methods for acoustic scene
  classification.
\newblock {\em Applied Sciences}, 10(6), 2020.

\bibitem{ruder2017overview}
Sebastian Ruder.
\newblock An overview of multi-task learning in deep neural networks.
\newblock {\em arXiv preprint arXiv:1706.05098}, 2017.

\bibitem{kingma2014adam}
Diederik~P Kingma and Jimmy Ba.
\newblock Adam: A method for stochastic optimization.
\newblock {\em arXiv preprint arXiv:1412.6980}, 2014.

\end{thebibliography}

\end{document}